\documentclass[twocolumn,twoside,slac_two]{revtex4}
\usepackage{graphicx}
\usepackage{fancyhdr}
\usepackage{hyperref}
\hypersetup{
    colorlinks=true,
    urlcolor=magenta,
}
\begin{document}

\title{Assessment of Source and Transport Parameters of Relativistic SEPs Based on Neutron Monitor Data}

%

\author{R. B\"{u}tikofer}
\affiliation{University of Bern, Physikalisches Institut, Sidlerstrasse 5, CH-3012 Bern, 
Switzerland}
\author{N. Agueda}
\affiliation{University of Barcelona, Institut de Ci{\`e}ncies del Cosmos, 
Facultat de F{\'i}sica Mart{\'i} i Franqu{\`e}s, 
1, E-08028 Barcelona, Spain}

\author{B. Heber, D. Galsdorf}
\affiliation{Christian-Albrechts-Universität zu Kiel, IEAP - Extraterrestrische Physik, Leibnizstr. 11, D-24098 Kiel, 
Germany}
\author{R. Vainio}
\affiliation{University of Turku, FI-20014 Turun Yliopisto, Finland}

\begin{abstract}
As part of the HESPERIA Horizon 2020 project, we developed a software package for the direct inversion of Ground
Level Enhancements (GLEs) based on data of the worldwide network of Neutron Monitors (NMs). The new methodology
to study the release processes of relativistic solar energetic particles (SEPs) makes use of several models, including: 
the propagation of relativistic SEPs from the Sun to the Earth, their transport in the Earth's magnetosphere and 
atmosphere, as well as the detection of the nucleon component of the secondary cosmic rays
by ground based NMs. The combination of these models allows to compute the expected 
ground-level NM counting rates caused by a series of 
instantaneous particle releases from the Sun. The proton release-time profile at the Sun 
and the interplanetary transport conditions are then inferred by fitting NM 
observations with modeled NM counting rates. In the paper the 
used models for the different processes, the software and first findings with the new software are presented.
\end{abstract}

\maketitle

\thispagestyle{fancy}

\section{INTRODUCTION}
The worldwide network of neutron monitors (NMs) together with the Earth's magnetic 
field acts as a huge spectrometer for cosmic rays (CRs) in the energy range $\sim$500 MeV to 
$\sim$15 GeV. The NM measurements are particularly useful for the quantitative 
investigations of energetic solar cosmic ray (SCR) events, so-called Ground Level 
Enhancements (GLEs). 
During the last decades NM data during GLEs were used to assess the characteristics 
of SCR near Earth, i.e.\ the transport in the interplanetary space was not 
included in these analyses and thereby no information about the SCR characteristics at the 
solar source could be derived. 

As part of the HESPERIA Horizon 2020 project a new approach was developed, i.e.\ a software package for the direct 
inversion of GLEs based on NM data. With the new GLE inversion software it is possible to directly assess the 
release timescales of relativistic SEPs at the Sun and the characteristics of 
their transport in interplanetary space based on NM observations of the worldwide 
network. 
Goals of the new GLE inversion software are to learn about the high-energy processes that release high energy 
($>$500 MeV) protons at the Sun and about the processes that affect their propagation in 
the interplanetary space up to the Earth orbit. 

\begin{figure}
\begin{center}
\includegraphics[width=58mm]{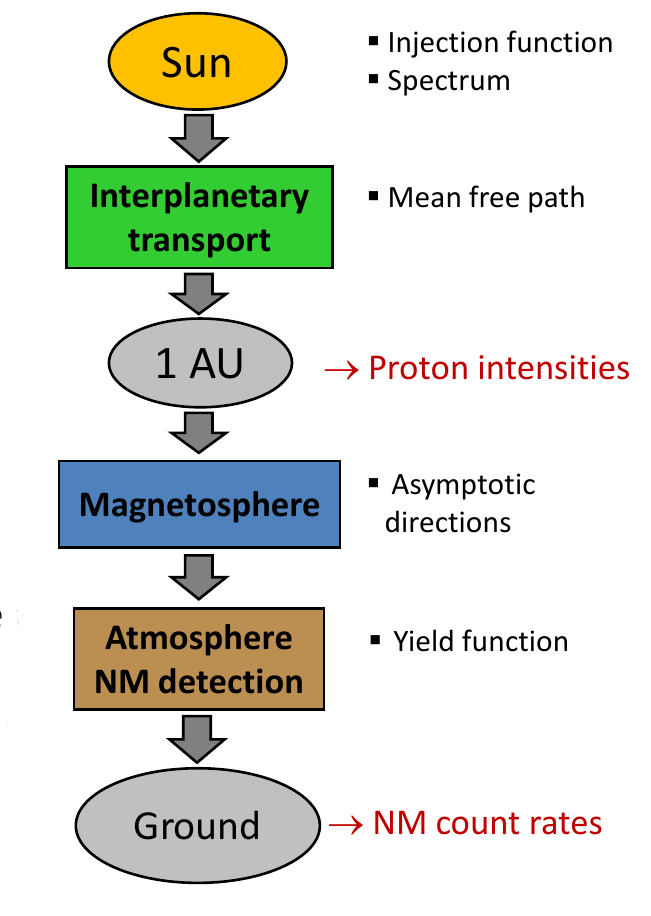}
\end{center}
\vspace*{-0.5cm}
\caption{Computation of the NM response to a particle release at the Sun.}
\label{schema}
\end{figure}

The procedure uses several models: propagation of relativistic SEPs in the interplanetary space from the Sun to the 
Earth, transport in the geomagnetosphere and in the Earth's atmosphere as well as the detection of the nucleon 
component of the secondary CRs by the NMs. The concept is summarised in Fig.~\ref{schema}.

\section{TRANSPORT FROM THE SUN TO THE EARTH}

\subsection{Injection of SEPs at the Sun and interplanetary transport of relativistic solar protons}
The injection profile at the Sun is described by a superposition of impulsive (delta) injections of particles at a given 
starting time. 
It is assumed that the solar source spectrum of the released solar protons has the form of a power law in rigidity with 
spectral index $\gamma$ ($dN/dR \propto R^{-\gamma}$).
The duration and the dependence of the particle release process as well as $\gamma$ is determined in the inversion 
procedure. 

\begin{figure}
\includegraphics[width=82mm]{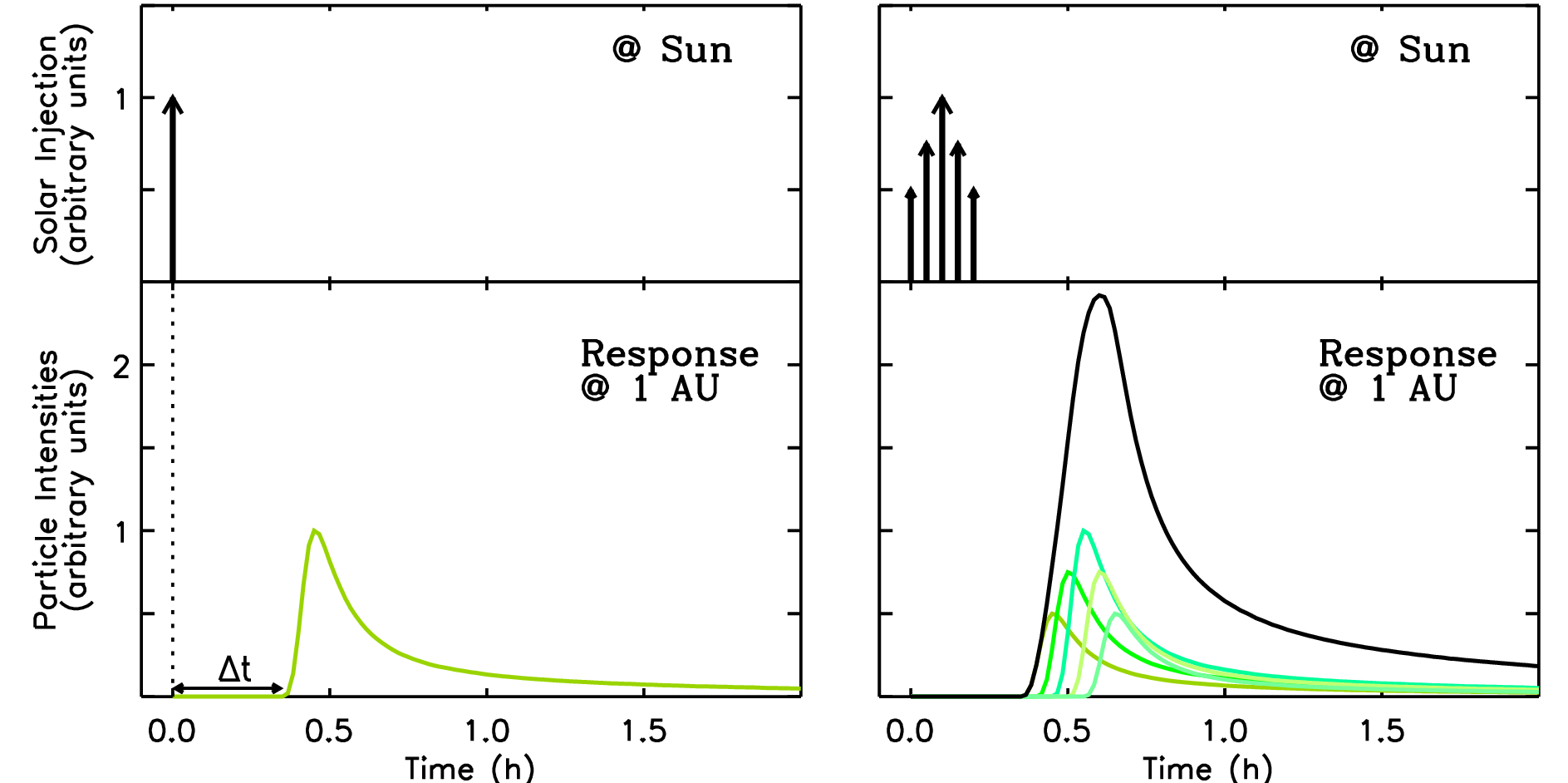}
\caption{Impulsive injection (delta function) at the Sun and the resulting response at 1~AU (left) and the 
superposition of impulsive injections and the corresponding response 
    at 1~AU, i.e.\ the sum of series of impulse responses.\label{convolution}}
\end{figure}

The interplanetary transport models provide the response of the system to an impulsive (delta) injection at the Sun, 
i.e.\ the Green's function of particle transport. Figure~\ref{convolution} shows an example of the output of 
the system (i.e.\ particle intensities at 1~AU from the Sun) for a well defined input such as a delta function at the 
Sun (left panel) as well as the superposition of impulsive injections (right panel).

In an unperturbed solar wind, the interplanetary magnetic field (IMF) can be described as a smooth average field, 
represented by an Archimedean spiral, with superposed magnetic fluctuations. In this case, the propagation of SEPs along 
the IMF is affected by both the effects of adiabatic motion along the smooth field and pitch angle scattering by 
magnetic turbulences. The quantitative treatment of the evolution of the particle's phase space density, $f(z,\mu,t)$, 
can be described by the focused transport equation \citep{Roelof69}:

\begin{equation}
    \frac{\partial f}{\partial t}+v \mu\frac{\partial f}{\partial z}+\frac{1-\mu^2}{2L}v \frac{\partial f}{\partial 
\mu}-\frac{\partial}{\partial \mu}\left(D_{\mu\mu} 
\frac{\partial f}{\partial \mu}\right)=q(z,\mu,t)\label{focutransp}
\end{equation}

Here $z$ is the distance along the magnetic field line, $\mu = \cos \alpha$ is the cosine 
of the particle pitch angle 
$\alpha$, and $t$ is the time. The systematic force is characterized by the focusing length, $L(z) = -B(z)/(\partial 
B/\partial z)$, in the diverging magnetic field $B$, while the stochastic forces are described by the pitch angle 
diffusion coefficient $D_{\mu\mu}(\mu)$. The injection of particles close to the Sun is given by $q(z,\mu,t)$. 
This simple form of the transport equation neglects convection and adiabatic deceleration 
(see \citep{Ruffolo95} for the full equation),
but for relativistic protons, this is typically a small effect. The 
effects of diffusion 
perpendicular to the average magnetic field and drifts \citep{DallaEtAl13} are also neglected. 
Instead, it is assumed that there is no variation across the magnetic field, and that the respective solutions are 
identical in neighboring flux tubes. 
A further limitation of the software in the present form is that it is not usable for the study of events that show 
bidirectional pitch angle distribution.

\begin{figure}
\includegraphics[width=75mm]{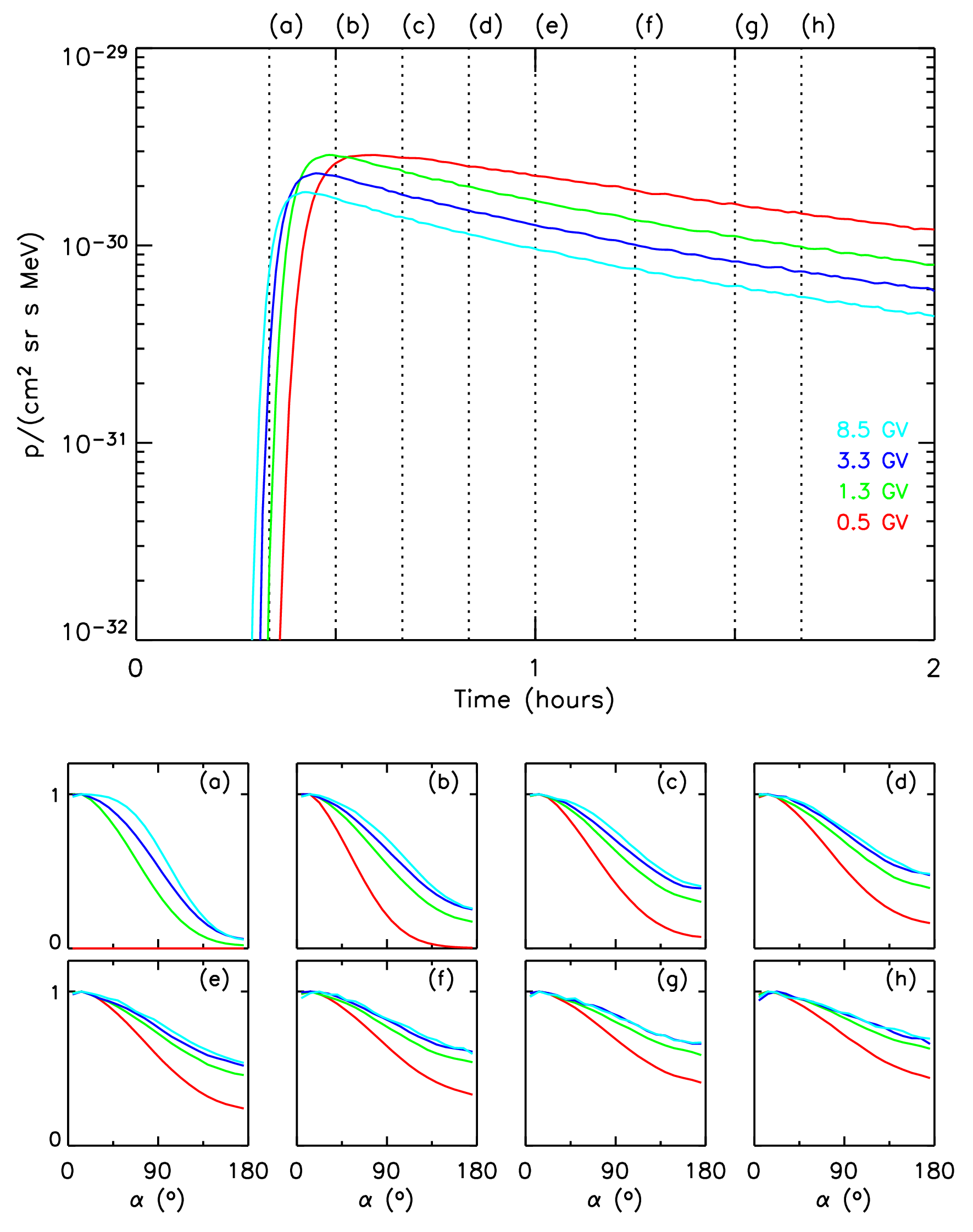}
\caption{Top: Proton omni-directional intensities at 1~AU for selected rigidities 
assuming $v_{sw}$ = 400~km s$^{-1}$, $\lambda_0$ = 0.2~AU and $\gamma$ = 3. 
Bottom: Corresponding pitch angle distributions for selected rigidities at eight 
different times (marked by the vertical lines in the top plot).}
\label{response}
\end{figure}

Analytical solutions of the focused transport equation are not known and numerical methods are applied to solve the 
focused transport equation. A Monte Carlo transport model \citep{AguedaEtAl08} is used to compute the 
proton intensities expected at 1~AU assuming that the solar source is static at two solar 
radii, the particle release is instantaneous and that particles are hypothetically moving 
at the speed of light. 
These Green's functions of particle transport for $v = c$ can be scaled to provide the Green's 
functions for any other mono-energetic particles. 
As an example Fig.~\ref{response} shows the omni-directional intensities and pitch angle distributions expected at 1~AU 
for several rigidity values, assuming a solar wind speed of 400~km s$^{-1}$, a mean free path for pitch angle scattering 
$\lambda_0$  = 0.2 AU and a spectral index $\gamma$ = 3 for the rigidity spectrum at the source.

\subsection{Transport in the Earth's magnetosphere}
In the geomagnetosphere, where the magnetic field (25-65~$\mu$T at the Earth's surface) is stronger than in the 
interplanetary space (typical 5~nT), CR particles are stronger deflected. The CR particle 
trajectories in the Earth's magnetic field are computed by numerical integration of backward trajectories in a model of 
the geomagnetic field, see e.g.\ \citep{Smart2000}. Thereby the effect is used that the path of a negatively charged 
particle with mass, $m$, charge, $q$, and speed, $v$, in a static magnetic field, $B$, is identical to that of a 
positively charged particle but with reverse sign of the velocity vector. Therefore, the common method to compute 
CR trajectories in the geomagnetic field is to calculate the trajectory in the reverse direction, i.e.\ the starting 
point of the reverse trajectory calculation is above the observation location in question at an altitude of typically 20 
km asl.

The asymptotic directions for all NM stations and for all historic GLEs, from GLE 1 through 71, were 
computed every 15~minutes using the software suite PLANETOCOSMICS~\citep{PLANETOCOSMICS}. The geomagnetic field is 
described by the IGRF model for the magnetic field contributions caused by sources in the Earth's interior and the 
Tsyganenko 1989 model~\citep{Tsy89}  describing the magnetic field caused by current systems within the 
geomagnetosphere.

\subsection{Transport in the atmosphere and the detection of secondary cosmic rays by the neutron monitors}
The transport of CR particles through the Earth's atmosphere and the detection of 
the nucleonic component of the secondary CRs are combined in the so-called NM yield function. Thereby
the NM yield function can directly be used to determine the CR flux at the top of the Earth's
atmosphere from the worldwide measurements of NMs.

In the current version of the GLE inversion software only the NM yield function for primary protons as determined by 
\citep{2008ICRC....1..289F} is integrated. This yield function is based on Monte Carlo 
calculations with Geant4~\citep{Agostinelli_NIMA_2003} for the transport in the Earth's 
atmosphere as well as for the efficiency of the NM for nucleons. 

The comparison of the results of different GLE analysis for a single GLE in the past have 
shown that the results may differ considerably~\citep{BuetikoferFlueckiger2015}. 
The main reason for this behavior may be that different yield functions were used.
Hence it is planned that also yield functions as published by other 
authors will be integrated in the GLE inversion software in the future.

\section{Software}
The GLE inversion software suite, written in the programming language ``Python'', includes several stand-alone modules. 
To run this software, the user needs to consider two main blocks. The first one (Modules 1-3) concerns the storage of 
the data needed for the GLE inversion. 
The second block (Module 4) is the actual GLE inversion.

\begin{description}
 \item[Module 1:] To provide access to NM data using the NEST tool under the web page of  
NMDB (Neutron Monitor Data Base, http://www.nmdb.eu), allowing a direct retrieval from 
NMDB to the user’s computer.
 \item[Module 2:] To download the asymptotic directions of NM stations for the selected time of the GLE under 
investigation from the HESPERIA webpage.
 \item[Module 3:] To download OMNI magnetic field data in GSE coordinates and perform time averages to get an 
estimation of the predetermined axis of symmetry SCR angular distributions. 
This magnetic field direction is used to compute the pitch angles observed by each NM 
station, i.e.\ the angles between the magnetic field direction and 
the asymptotic directions of the NM station.  

\item[Module 4:] To compute the count rate increases for each selected NM station based 
on the modeled
intensities of CR particles and to perform the inversion using the data recorded by the selected NMs and the 
simulated count rates.
\end{description}

The following parameters are determined by the GLE inversion software: the series of instantaneous releases from the 
Sun, the source spectral index, $\gamma$, and the mean free path, $\lambda_0$.

\section{Case study: GLE on 15 April 2001}
\begin{figure}
\includegraphics[width=58mm]{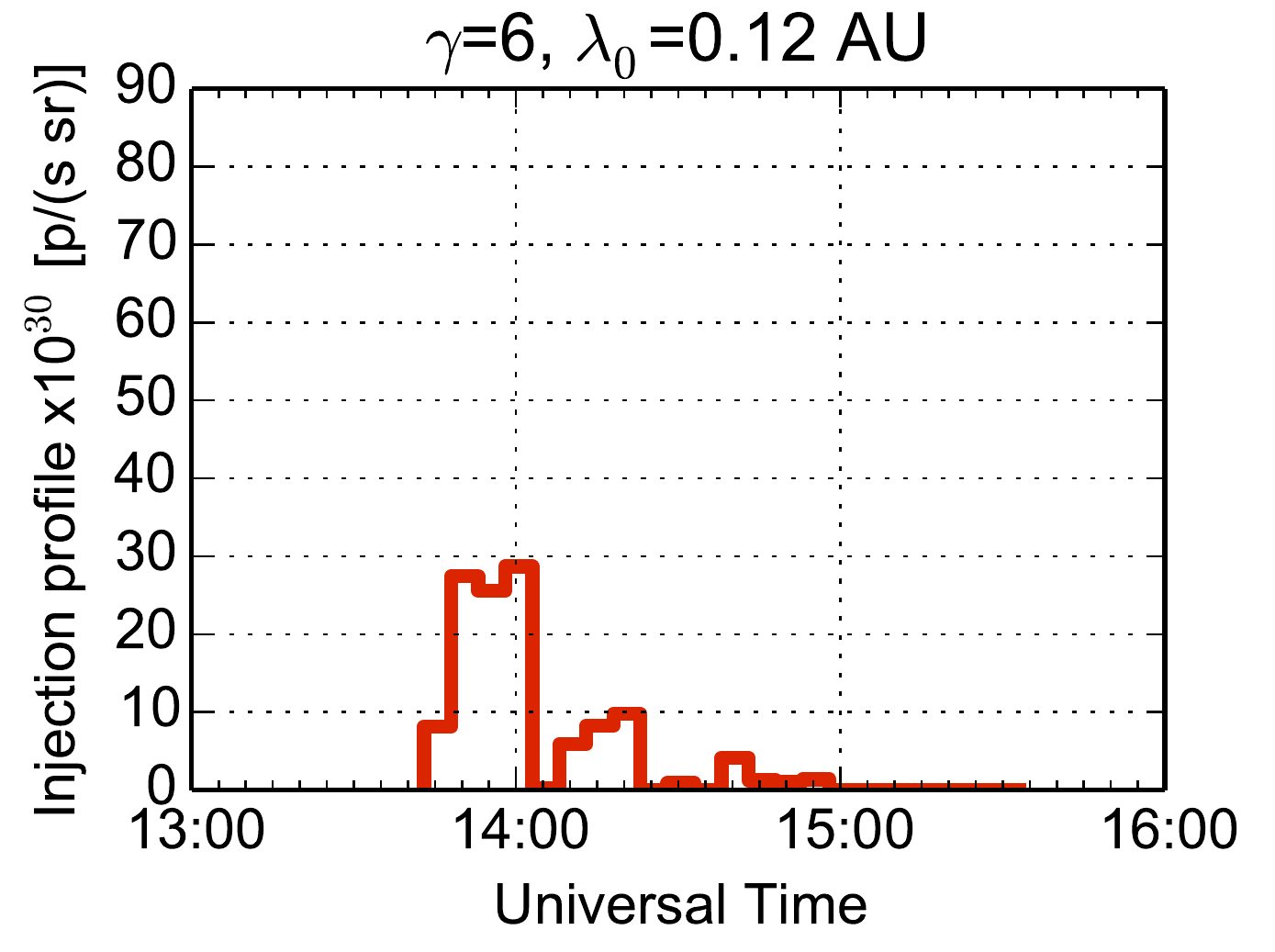}
\vspace*{-0.2cm}
\caption{Series of instantaneous releases at the Sun during the GLE on 15 April 2001 that best fits the measured data 
of the selected NMs with $\lambda_0 = 0.12$~AU and $\gamma = 6$ when applying the GLE inversion software introduced in 
this work.}
\label{GLE60_injection_profile}
\end{figure}

\begin{figure}
\hspace*{-0.08cm}
\includegraphics[width=77mm]{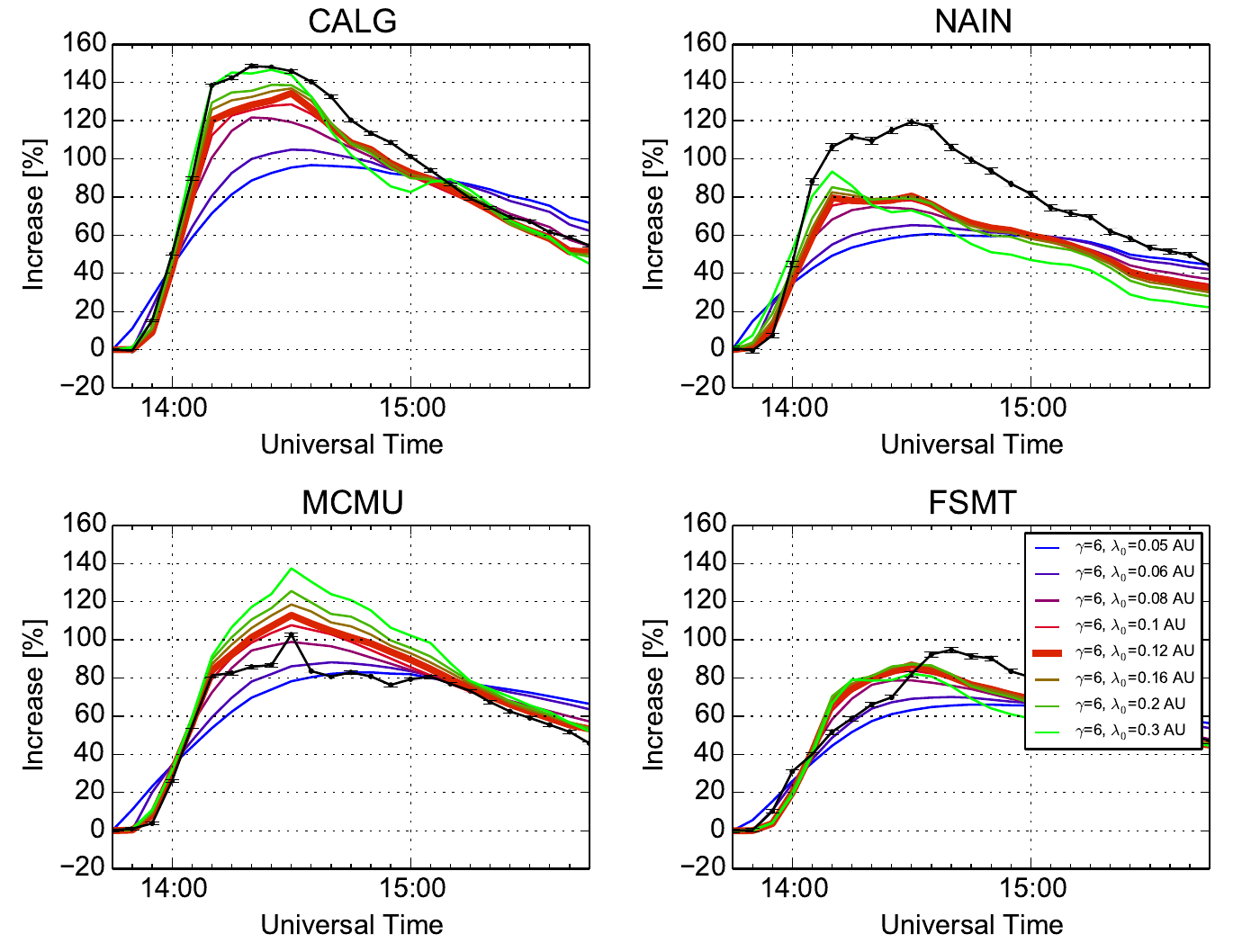}
\vspace*{-0.10cm}
\caption{Relative count rate increase during GLE on 15 April 2001 as measured by the NM stations with the largest 
observed count rate increases (Calgary, Nain, McMurdo, and Fort Smith, black curve) and as simulated for $\gamma= 6.0$ 
and different mean free path lengths for pitch angle scattering $\lambda_0$ (colored curves). The red bold curve 
corresponds to the best fit, i.e.\ injection profile as shown in Fig.~\ref{GLE60_injection_profile}, $\gamma= 6.0$, and 
$\lambda_0 = 0.12$~AU.}
\label{GLE60_sim_vs_measured_NM}
\end{figure}

As a first case study the GLE inversion software was applied during the GLE on 15 April 2001 (GLE60). 
Figure~\ref{GLE60_injection_profile} shows the injection profile at the Sun, the mean free path for scattering in the 
interplanetary space $\lambda_0 = 0.12$~AU, and power law index $\gamma = 6$ for rigidity spectrum at the source. 
These characteristics best fit the measured data of the selected NM stations Apatity, Calgary, Fort Smith, Irkutsk, 
Kerguelen, Kiel, Nain, McMurdo, Oulu, Rome, Terre Adelie, and Tixie Bay in the time interval 1345--1545~UT.
In Fig.~\ref{GLE60_sim_vs_measured_NM} the simulated and measured relative count rate increases for the NM stations with 
the highest count rate increases are presented, i.e.\ with the injection profile shown in 
Fig.~\ref{GLE60_injection_profile}, $\gamma = 6$ and selected values for $\lambda_0$.

\section{Summary}
Within the HORIZON~2020 project HESPERIA a new GLE analysis software was developed. Compared to former GLE inversion 
computer programs, which assessed the SCR characteristics near Earth, the current software determines the SCR 
characteristics at the Sun in one step, i.e.\ during the whole GLE (typically over one 
hour) as well as the characteristics of the transport between the Sun and 1~AU. 
There are made different assumptions and simplifications in the particle release at 
the Sun as well as in the transport in the interplanetary space. In addition, the 
transport in the Earth's magnetic field is based on magnetic field models which may show 
inaccuracies, and finally the used yield function differs somewhat from other published 
NM yield functions.    

A first case study with the GLE on 15 April 2001 gives confident results. However, the GLE 
inversion software must be 
applied during further GLEs, e.g.\ GLEs with smaller count rate increases at the NM stations, GLEs with more 
complicate conditions in the interplanetary space or in the release proportion at the Sun. The software is available to 
the community from the HESPERIA webpage 
(http://www.hesperia-space.eu/index.php/results/inversion-software-tools).

\bigskip 
\begin{acknowledgments}
This project has received funding from the European Union’s Horizon 2020 research and innovation programme under grant 
agreement No 637324. The work was supported by the Spanish Project AYA2013-42614-P and the Swiss State Secretariat for 
Education, Research and Innovation (SERI) under the contract number 15.0233. 
We thank the NM data base NMDB and the investigators of the following NM stations for the data that we used for this 
analysis: Apatity, Calgary, Fort Smith, Irkutsk, Kerguelen, Kiel, Nain, McMurdo, Oulu, Rome, Terre 
Adelie, Tixie Bay.
\end{acknowledgments}

\bigskip 

\begin{thebibliography}{99}   

\bibitem{Roelof69}
E.C. Roelof, 
in Lectures in High-Energy Astrophysics, 1969.

\bibitem{Ruffolo95}
D. Ruffolo, 
Astrophysical Journal, 442, 861, 1995.

\bibitem{DallaEtAl13}
S. Dalla, M.S. Marsh, J. Kelly, T. Laitinen, 
Journal of Geophysical Research, 118, 5979,2013.

\bibitem{AguedaEtAl08}
N. Agueda, R. Vainio, D. Lario, B. Sanahuja, 
Astrophysical Journal, 675, 1601, 2008.

\bibitem{Smart2000}
D.F. Smart, M.A. Shea, E.O. Fl{\"u}ckiger, 
Space Sci. Rev., 93, 305, 2000.

\bibitem{PLANETOCOSMICS}
 L. Desorgher, 
http://cosray.unibe.ch/\textasciitilde laurent/planetocosmics, 2005.

\bibitem{Tsy89}
N. A. Tsyganenko, 
Planet. Space Sci. 37, 5-20, 1989. 

\bibitem{2008ICRC....1..289F}
E.O. Fl{\"u}ckiger, M.R. Moser, B. Pirard, et al., 
in Proceedings of 30th International Cosmic Ray Conference, 1, 289,  2008.

\bibitem{Agostinelli_NIMA_2003}
S.Agostinelli et al., 
Nuclear Instruments and Methods in Physics Research A, 
506,250, 
2003.

\bibitem{BuetikoferFlueckiger2015}
R. B{\"u}tikofer, E. Fl{\"u}ckiger, 
Journal of Physics Conference Series, 632(1), 012053, 2015. 
\end{thebibliography}

\end{document}